\documentclass[prl,superscriptaddress,a4paper,floatfix,nofootinbib,longbibliography]{revtex4-1}
\usepackage{amsmath}
\usepackage{float}
\usepackage{amsfonts}
\usepackage{fancybox}
\usepackage{eepic}
\usepackage{xcolor}
\usepackage{times}
\usepackage{latexsym}
\usepackage{pifont}
\usepackage{graphicx}
\usepackage{epstopdf}
\usepackage{bm}
\usepackage{scrextend}
\usepackage{amsfonts}

\def\kk{\boldsymbol{k}}

\newcommand{\qsgw}{QS$GW$}
\newcommand{\qsgwl}{QS$G\hat{W}$}
\newcommand{\qub}{NI-HPC, Queen's University Belfast, Belfast BT7 1NN, Northern Ireland, United Kingdom}

\begin{document}
\title{Many-Body Vertex Effects: Time-Dependent Interaction Kernel with Correlated Multi-Excitons in the Bethe-Salpeter Equation}
\author{Brian Cunningham}
\affiliation{\qub}

\begin{abstract}
Building on a beyond-$GW$ many-body framework that incorporates higher-order vertex effects in the self-energy \textemdash giving rise to $T$-matrix and second-order exchange contributions \textemdash this approach is extended to now include the vertex derived in that work to the kernel in the Bethe-Salpeter Equation (BSE) for the reducible polarization function. This results in a frequency-dependent interaction kernel that naturally captures random phase approximation (RPA) effects, dynamical excitonic interactions, and the correlated propagation of multiple correlated electron-hole pairs that model multi- (including bi- and tri-) excitonic effects, relevant for nonlinear optics and high harmonic generation. These processes emerge from including the functional derivatives of the screening and vertex with respect to the Green’s function in the vertex, enabling a fully $abinitio$, time-dependent treatment of correlation effects. By focusing on the reducible rather than irreducible polarization function, this approach provides a computationally viable framework for capturing complex many-body interactions for calculating the self-energy, optical spectra and EELS. The resulting interaction kernel is relatively straightforward, clearly delineates the physical processes that are included and omitted, and has the same dimensionality as the conventional BSE kernel used in standard many-body theory implementations, but is now itself frequency dependent. The method is expected to facilitate the integration of advanced many-body effects into state-of-the-art software packages, offering a universal and highly accurate framework for the description of sub-atomic correlations. Such advancements are crucial for the development of semiconductor, optoelectronic, superconducting and antimatter technologies, and ensuring that theoretical modeling evolves alongside exascale and accelerated computing.

\end{abstract}
\maketitle

%
%
%
\section{Introduction}
Advancing our understanding of sub-atomic particle interactions \textemdash~such as electron-electron, electron-nucleus, and electron-positron interactions \textemdash~is essential for increasing our understanding of the universe, and driving progress in technology, medicine and computation. The Kohn-Sham formulation of density functional theory (DFT)\cite{PhysRev.136.B864,kohnsham} is the most widely used method for studying electronic interactions due to its balance of accuracy and efficiency. DFT continues to evolve, being applied to real-world problems, however, despite the success of DFT and its extensions, such as generalized Kohn-Sham schemes\cite{PhysRevB.53.3764}, DFT+U,\cite{PhysRevB.44.943} and DFT+DMFT,\cite{PhysRevB.82.045105} its limitations in strongly correlated systems necessitate the development of more accurate many-body approaches
\par
Whilst it is important we continue to further refine and extend DFT-based methods, it is equally important to develop higher-accuracy approaches, such as wavefunction methods\cite{CI_GPU} and many-body theory.\cite{hedin} DFT struggles when applied to strongly correlated systems such as NiO.\cite{ferdi94,QSGWhat,NiODFT} Recent rapid advances in exascale and GPU computing\cite{molGW,CI_GPU} suggest that computationally expensive methods \textemdash~approaching current practical limits \textemdash~may become more accessible in the near future. Wavefunction methods, while highly accurate, are often restricted to the calculation of ground-state properties in smaller systems, requiring significant effort to extract additional information, that quite often is not of any use\cite{kohn_lecture}. 
\par
To address these challenges, many-body theory (MBT) approaches, particularly those based on Lars Hedin’s formalism,\cite{hedin} provide a promising balance between accuracy and computational efficiency. A real benefit of the approach is that it is \textit{abinitio} and can be extended with the addition of more diagrams describing processes missing from a particular implementation\cite{cunningham24}. MBT also has the benefit that excited state phenomena (required for investigating, e.g., optoelectronic phenomena) are readily computed. Most MBT implementations are based on the $GW$ approximation\cite{louie,GW_aryasetiawan,hedinGW} to Hedin's equations. Whilst an improvement over DFT, many implementations are simply a first-order perturbation applied to the DFT electronic structure, and as a result depend on the DFT functional chosen, and therefore tend to perform poorly in strongly correlated systems.\cite{sabatino23,QSGW_paper} Another key issue with $GW$ is the inadvertent cancellation of errors due to missing diagrams, which can lead to seemingly accurate results for specific quantities, such as band gaps, while failing to capture other important physical phenomena, and quantities such as the total energy.\cite{Miyake02,doi:10.1021/acs.jctc.3c01200} Although this may prove beneficial in certain circumstances, it is important we do not overlook this, as certain physical processes may not be captured at all, limiting the method's predictive reliability in unexplored materials. Self-consistent approaches remove the starting-point dependence by iterating the $GW$ formalism, however, these can also prove problematic.\cite{SC_ener,ferdi94} The quasiparticle self-consistent $GW$ (QS$GW$) method\cite{QSGW_paper,QSGWhat,QSGW_prl1,QSGW_PRL} is one such method that has become popular in recent years. However, conventional QS$GW$ ignores vertex effects in both the polarization function and self-energy and as a result performs worse in simple systems due to one competing effect now being included and no longer cancelling with another missing effect. 
\par
To address these shortcomings, researchers have exploreed incorporating vertex effects in the polarization function and/or self-energy. Inclusion of vertex effects in the polarization function through solving the Bethe-Salpeter equation (BSE) has become popular in recent years\cite{PhysRevLett.91.056402,PhysRevLett.91.256402,yambo}. Solving the BSE with an effective non-local static kernel derived from time-dependent DFT\cite{PhysRevB.81.085213,PhysRevLett.99.246403,PhysRevLett.94.186402} has become a common approach. Another approach is to solve the BSE with the kernel $i\delta\Sigma/\delta G$ (the functional derivative of the self-energy with respect to the Green's function) calculated with the $GW$ approximation for $\Sigma$, whilst neglecting $\delta W/\delta G$ and assuming a static kernel. We used the latter for the QS$G\widetilde{W}$ method in the Questaal code\cite{opt_PRM,QSGWhat} to capture excitonic effects in strongly correlated systems such as NiO for optical spectra and to also improve the electronic structure. In those works we calculated $W$ at the level of the random phase approximation (RPA). Whilst an improvement on QS$GW$ there are still many effects missing from the method, including time-dependent interactions in the BSE and multiexcitonic effects. Time-dependence in the interaction kernel is examined in Ref.~\cite{PhysRevLett.91.176402,williams21} and shown to strongly affect optical spectra in noble metals.\cite{PhysRevLett.91.176402} Multiexcitonic effects\cite{multicc,PhysRevB.93.140409} describe the process of a number of interacting electron-hole pairs propagating in the system. These effects are crucial for examining optoelectronic phenomena such as high-harmonic generation\cite{hhg} and non-linear effects. 
\par
Another effect mssing in the QS$G\widetilde{W}$ method is the vertex in the expression for the self-energy, i.e. going beyond the $GW$ approximation for $\Sigma$. Methods exist for includng vertex effects in the self-energy, such as those developed by Kutepov\cite{PhysRevB.94.155101,PhysRevB.95.195120}, showing the significance of these effects in predicting flat, dispersionless bands in materials like the copper halides CuCl and CuBr.\cite{QSGWhat,kutepov_cucl} The multichannel Dyson equation method\cite{PhysRevLett.131.216401} and the method outlined in Ref.~\cite{PhysRevB.106.165129} also included vertex effects in the self-energy. Recently, the author developed a method for including vertex effects in the self-energy,\cite{cunningham24} yielding a relatively straightforward expression for the self-energy, that extends $GW$ by incorporating $T$-matrix diagrams and second-order statically screened exchange. In that work the benefit of working with the reducible polarization function instead of the irreducible one was made clear. We also examined a similar method with the inclusion of the $T$-matrix diagrams applied to the case of a positron interacting in finite systems,\cite{nature_positron} with the additions proving essential for capturing positron physics (describing processes such as virtual positronium formation) and accurately predicting positron binding energies\cite{nature_positron} and scattering and annihilation rates.\cite{prl_pos}
\par 
\par
In this work, the beyond-$GW$ formalism ($GW\Gamma$) introduced in Ref.~\cite{cunningham24}, that derives an analytic expression for the interaction kernel $i\delta\Sigma/\delta G$ from Hedin’s equations, is extended to examine these same effects in the polarization function. This approach incorporates time-dependent effects and multiexcitonic interactions, addressing key limitations of conventional BSE formulations. By including the functional derivative of the vertex from $\Sigma=iGW\Gamma$, correlations between seperate electron-hole pairs, including higher-order exchange effects, arise. These new additions give rise to effects beyond the standard BSE, including multiexcitonic effects. The derivation is performed in the time domain before being transformed to frequency, and the key steps are presented for the possibility of examining, e.g., time-dependent interactions in the $T$-matrix and second-order screened exchange diagrams in the future. The method relies on using the reducible polarization function $\Pi$ instead of the irreducible one, as the screening is simply $v+v\Pi v$, instead of $(1-vP)^{-1}v$ for the irreducible one. The final expression is depicted with diagrams illustrating the processes included in the formalism. The result is a matrix equation for the reducible polarization function akin to the usual BSE, but now with a frequency dependent interaction kernel that contains higher order diagrams. The suppression of self-polarization effects (more pronounced in strongly correlated systems) emerges naturally in this framework. The approach is general, intuitive and does not rely on empirical or ad-hoc parameters/interactions and when combined with Ref.~\cite{cunningham24} is fully self-contained, having the potential to be implemented in various many-body electronic structure and even antimatter\cite{nature_positron} software packages.
\par
\section{Derivation of the Polarization function}
The single particle Green's function in the frequency domain is
\begin{equation}
G(\boldsymbol{r},\boldsymbol{r}',\omega)=\sum_{n}\frac{\varphi_n(\boldsymbol{r})\varphi_n^{*}(\boldsymbol{r}')}{\omega-\varepsilon_n\pm i\eta},
\end{equation} 
where $\varphi_n$ and $\epsilon_n$ are the eigenfunctions and eigenvalues of the corresponding Hamiltonian, $\eta$ is a positive infinitesimal and the $+(-)$ is for particles(holes). $n$ may be a composite index subsuming spin and orbital index, or band index and $\boldsymbol{k}$ in extended systems. The inverse Fourier transform $\frac{1}{2\pi}\int G(\boldsymbol{r},\boldsymbol{r}',\omega)e^{-i\omega t}d\omega$, produces
\begin{equation}
G(\boldsymbol{r},\boldsymbol{r}',t)=\sum_n\varphi_n(\boldsymbol{r})\varphi_n^{*}(\boldsymbol{r}')G_n(t),
\end{equation}
where $G_n(t)=\mp ie^{-i\varepsilon_n t}e^{\mp\eta t}\theta(\pm t)$ and the transformation is calculated using the Cauchy residue theorem 
\begin{equation}
\oint\limits_{C}f(z)dz=\pm 2\pi i\sum{\rm residues~inside~C}, 
\end{equation}
with the contour $C$ taken in the upper plane for $t<0$ (holes) and in the lower plane for $t>0$ (particles) to ensure convergence of $e^{\mp\eta t}$ as $|t|\rightarrow \infty$. This will be useful for determining the interaction kernel later. The Green's function is diagonal in the basis of the single particle states $\varphi_n(\boldsymbol{r})$.

The basis of particle-hole states is used throughout this work for quantities involving two particles,
\begin{eqnarray}
S_{\scriptsize\begin{array}{l}n_1n_2\\n_3n_4\end{array}}=\int\int\int\int d\bm{r}_1d\bm{r}_2d\bm{r}_3d\bm{r}_4\varphi_{n_1}^{*}(\bm{r}_1)\varphi_{n_2}(\bm{r}_2)S(\bm{r}_1,\bm{r}_2,\bm{r}_3,\bm{r}_4)\varphi_{n_3}(\bm{r}_3)\varphi_{n_4}^{*}(\bm{r}_4),
\end{eqnarray}
where $S$ is a function of four space cordinates and the $n$ index the eigenstates of the single particle Hamiltonian, denoting a particle (unoccupied ground state) or a hole (occupied ground state). $S$ can also be a two or three point function by making use of the Dirac delta function, e.g., the bare Coulomb interaction is $v(\bm{r}_1,\bm{r}_2,\bm{r}_3,\bm{r}_4)=v(\bm{r}_1,\bm{r}_3)\delta(\bm{r}_1-\bm{r}_2)\delta(\bm{r}_3-\bm{r}_4)$. The transformation back to real space can be performed using the fact the single particle states form a complete orthonormal basis. 
From the Hedin equation for the irreducible polarization function\cite{cunningham24,hedin,onida_electronic_2002} $P(1,2)=-iG(1,3)G(4,1)\Gamma(3,4;2)$\footnote{Equations for the self-energy and polarization function are often written with an infinitesimal time shift $t=t+\eta$ (e.g., $G(1,2^{+})$ to clarify the time-ordering at equal times and avoid ambiguities when convolving with instantaneous interactions like the Coulomb interaction). This notational detail is omitted here, but the standard convention that such shifts are implicitly understood when evaluating expressions involving equal-time singularities are adopted.} (with $1=(\boldsymbol{r}_1,t_1,\sigma_1)$, i.e., space-time and possibly spin, and repeated indices on one side of the equality implying summation/integration) the zeroth order term, i.e., with the vertex approximated as $\Gamma(3,4;2)=\delta(3,4)\delta(4,2)$, in the basis of particle-hole states is
\begin{equation}
P^0_{\scriptsize\begin{array}{l}n_1n_2\\n_3n_4\end{array}}(\omega)=\frac{f_{n_2}-f_{n_1}}{\omega-\varepsilon_{n_1}+\varepsilon_{n_2}+(f_{n_2}-f_{n_1})i\eta}\delta_{n_1,n_3}\delta_{n_2,n_4}
\end{equation}
with $f$ the single particle occupations, 0(1) for a particle(hole), also denoted $c(v)$ for conduction(valence). Note that $n_1n_2$ denoting a particle-hole pair will result in a $+i\eta$ shift, i.e., a pair propagating forwards in time, and $n_1n_2$ denoting a hole-particle pair will result in a $-i\eta$ shift, representing a negative energy pair propagating backwards in time. In the time domain this is
\begin{equation}
P^0_{\scriptsize\begin{array}{l}n_1n_2\\n_3n_4\end{array}}(t)=-ie^{-i(\varepsilon_{n_1}-\varepsilon_{n_2})t}e^{\mp\eta t}\theta(\pm t)\delta_{n_1,n_3}\delta_{n_2,n_4},
\end{equation}
where the upper(lower) sign is for $n_1,n_2$ a particle-hole(hole-particle) pair. This represents the forward propagating non-correlated electron-hole propagator (or backward travelling hole-particle propagator). 
Let's now take Hedin's equations for the polarization function and many-body vertex,\cite{hedin,onida_electronic_2002}
\begin{equation}
\Gamma(1,2;3)=\delta(1,2)\delta(2,3)+\frac{\delta\Sigma(1,2)}{\delta G(4,5)}G(4,6)G(7,5)\Gamma(6,7;2),
\end{equation}
with $\Sigma$ the self-energy. As discussed in Ref.~\cite{cunningham24} the reducible polarization function $\Pi=P+Pv\Pi$, where $v(\boldsymbol{r},\boldsymbol{r}')=1/|\boldsymbol{r}-\boldsymbol{r}'|$ is the bare Coulomb potential, can replace the irreducible one, with  
\begin{equation}
\Pi(1,2)=P^0(1,2)+P^0(1;3,4)K(3,4;5,6)\Pi(5,6;2),\label{eq:PiBSE}
\end{equation} 
and the 3-point definitions of the polarization functions are used\footnote{The four-point $P^0(1,2;3,4)=-iG(1,3)G(4,2)$ is defined by `opening the legs' from the definition of the two-point polarization function.}, e.g., $P(1,2;3)=-iG(1,4)G(5,2)\Gamma(4,5;3)$, and $K(3,4;5,6)=v(3,5)\delta(3,4)\delta(5,6)+i\delta\Sigma(3,4)/\delta G(5,6)$.

The functional derivative of the self-energy is presented in Ref.~\cite{cunningham24} and when considered explicitly in the self-energy, $\Sigma(1,2)=iG(1,3)W(4,1)\Gamma(3,2;4)$ (with $W=v+v\Pi v$ the screened interaction), gives rise to higher-order diagrams that include the second-order screened exchange and diagrams describing the infinite series of 2-body ladder interactions between the added particles (or holes as the result of particle removal) and the excited electrons or holes in the system. The expression for the functional derivative of the self-energy in Ref.~\cite{cunningham24} is derived by also considering now the functional derivative of the screening and vertex, that are usually neglected in the BSE. These diagrams will now be included in the expression for the reducible polarization function. To do so, an expression for $\Pi$ with a time-dependent interaction kernel will need to be derived. 
If only the first term in Eq.~(14) for $i\delta\Sigma/\delta G$ from Ref.~\cite{cunningham24} is taken (i.e., adopting the $GW$ approximation for $\Sigma$ and ignoring $\delta W/\delta G$) and also assumed static (the $\omega=0$ contribution) then the usual static Bethe Salpeter Equation (BSE)\cite{BSE_paper,PhysRevLett.91.056402,PhysRevLett.91.256402,QSGWhat} can be derived from Eq.~\ref{eq:PiBSE}. The screening matrix elements in the basis of particle-hole pairs are then
\begin{equation}
W_{\scriptsize\begin{array}{l}n_1n_2\\n_3n_4\end{array}}(\omega)=v_{\scriptsize\begin{array}{l}n_1n_2\\n_3n_4\end{array}}+\sum_{{\scriptsize\begin{array}{l}n_5n_6\\n_7n_8\end{array}}}v_{\scriptsize\begin{array}{l}n_1n_2\\n_5n_6\end{array}}\Pi_{\scriptsize\begin{array}{l}n_5n_6\\n_7n_8\end{array}}(\omega)v_{\scriptsize\begin{array}{l}n_7n_8\\n_3n_4\end{array}},
\end{equation}
and for the simple vertex $i\delta\Sigma(3,4)/\delta G(5,6)=-W(4,3)\delta(3,5)\delta(4,6)\delta(t_3-t_4)$,
\begin{equation}
\Pi_{\scriptsize\begin{array}{l}n_1n_2\\n_3n_4\end{array}}(\omega)=(f_{n_4}-f_{n_3})\sum_{\lambda}\frac{X_{n_1n_2,\lambda}X^{-1}_{\lambda,n_3n_4}}{\omega-E_\lambda+i(f_{n_4}-f_{n_3})\eta},
\end{equation}
where $X$ and $E$ are the eigenvectors and eigenvalues of the 2-particle matrix $H_{\scriptsize\begin{array}{l}n_1n_2\\n_3n_4\end{array}}=(\varepsilon_{n_1}-\varepsilon_{n_2})\delta_{n_1n_3}\delta_{n_2n_4}\pm K_{\scriptsize\begin{array}{l}n_1n_2\\n_3n_4\end{array}}$, with $+(-)$ for $n_1$ a particle(hole). In this case, $K_{\scriptsize\begin{array}{l}n_1n_2\\n_3n_4\end{array}}=v_{\scriptsize\begin{array}{l}n_1n_2\\n_3n_4\end{array}}-W_{\scriptsize\begin{array}{l}n_4n_2\\n_3n_1\end{array}}(\omega=0)$. Calculating the BSE for the reducible polarization function has the benefit of being able to calclate the poles of $W$ (exactly equal to the poles of $\Pi$) and to determine the $GW$ self-energy, we no longer need to construct and invert the dielectric matrix and perform numerical frequency integration (as in Ref.~\cite{QSGW_paper}), nor is, e.g., the plasmon-pole approximation required. Another benefit of calculating $\Pi$ is that it has the same structure as $P^0$ above and can be written in the time domain as the sum of a static part and terms describing forward and backward propagation, and so the screening matrix is then
\begin{equation}
\bm{W}(t)=\bm{v}\delta(t^+)+\bm{W}^{+}(t)e^{-\eta t}\theta(t)+\bm{W}^{-}(t)e^{+\eta t}\theta(-t),
\end{equation}
where $t^{+}=t+\eta$ and the forward ($+$) and backward ($-$) propagating (in time) parts of the correlation contribution to the screening are
\begin{equation}
W^{\pm}_{\scriptsize\begin{array}{l}n_1n_2\\n_3n_4\end{array}}\hspace{-0.15cm}(t)=- i\hspace{-0.1cm}\sum_{\lambda\scriptsize\begin{array}{l}ph(hp)\\n_5n_6\end{array}}\hspace{-0.15cm}v_{\scriptsize\begin{array}{l}n_1n_2\\n_5n_6\end{array}}\hspace{-0.15cm}X_{n_5n_6,\lambda}X^{-1}_{\lambda,ph(hp)}e^{- iE_\lambda t}v_{\scriptsize\begin{array}{l}ph(hp)\\n_3n_4\end{array}},
	\end{equation} with the upper(lower) sign corresponding to $t>(<)0$ and the summation restricted to particle-hole(hole-particle) pairs, depending on the sign of $t$. Note that to reduce memory requirements $vX$ can be calculated (or rather its two contributions) and transformed to an auxillary or reduced mixed product basis, as in Ref.~\cite{aux,QSGW_paper}. 

To include higher order terms in the kernel (as in Ref.~\cite{cunningham24}), a time-dependent interaction kernel is required. This also allows the incorporation of dynamical excitonic effects, as discussed in, e.g., Ref.~\cite{PhysRevLett.91.176402,williams21}. To do so, let's take Eq.~\ref{eq:PiBSE} and assume $t_3=t_5$ and $t_4=t_6$ and that the kernel is only dependent on a single time difference $t_3-t_4$, as in Ref.~\cite{cunningham24}, which discusses the diagrams missing from the vertex as a result. The polarization function is also assumed time translationally invariant, i.e., only dependent on the time difference $t_1-t_2$, and so one can arbitrarily set $t_2=0$ and the polarization function will depend solely on $t_1$. The Fourier transform is then performed on the expression for the polarization function from the time domain, with the four-point $P^0$ expressed in terms of the Green's functions. Begin by determining the first-order term in the expansion, $\Pi=P^0+P^0KP^0+...$, which can be extended to higher-order terms, whilst focusing only on the dynamical parts of the interactions for now,
	\begin{eqnarray}\Pi_{\scriptsize\begin{array}{l}n_1n_2\\n_3n_4\end{array}}^1(\omega)=(-i)^2\hspace{-0.12cm}\int\int\int\hspace{-0.12cm} dtdt_3dt_4 e^{i\omega t}G_{n_1}(t_1-t_3)G_{n_2}({t_4-t_1})K_{\scriptsize\begin{array}{l}n_1n_2\\n_3n_4\end{array}}(t_3-t_4)G_{n_3}(t_3-t_2)G_{n_4}(t_2-t_4),\end{eqnarray}
	with $t=t_1-t_2$. $n_1n_2$ and $n_3n_4$ will be particle-hole or hole-particle, but will be the same when the Tamm-Dancoff approximation\cite{TD_myrta,TDA2020} (TDA) is adopted. To evaluate, use the expression for the Heaviside step function \newline$\theta(t)=\lim_{\eta \rightarrow 0^+}1/2\pi i\int e^{ixt}/(x-i\eta)dx$ and make use of the relation $\int e^{i\omega t}dt=2\pi\delta(\omega)$, whilst making sure the contour is placed in the correct plane to ensure the poles are such that terms $e^{ixt}$ do not diverge for complex $x$ as $|t|\rightarrow\infty$. For $n_1n_2$ and $n_3n_4$ both particle-hole pairs,
	\begin{widetext}\begin{equation}
	\begin{array}{lr}\Pi^1_{\scriptsize\begin{array}{l}n_1n_2\\n_3n_4\end{array}}(\omega)=\frac{i}{\varepsilon_{n_1}-\varepsilon_{n_2}-\omega-i\eta}\int_{0}^{\infty}dt_3\int_0^{t_3}dt_4&\left\{K_{\scriptsize\begin{array}{l}n_1n_2\\n_3n_4\end{array}}(t_3-t_4)e^{i(\varepsilon_{n_2}-\varepsilon_{n_3}+\omega+i\eta)t_3}e^{i(\varepsilon_{n_4}-\varepsilon_{n_2}-i\eta)t_4}+\right.\\&\left.K_{\scriptsize\begin{array}{l}n_1n_2\\n_3n_4\end{array}}(t_4-t_3)e^{i(\varepsilon_{n_4}-\varepsilon_{n_1}+\omega+i\eta)t_3}e^{i(\varepsilon_{n_1}-\varepsilon_{n_3}-i\eta)t_4}\right\}\end{array}
	\end{equation}
	\end{widetext}
	where the terms arise from the two possible time orderings of the interaction. Upon changing the integration variable $t=t_3-t_4$ we get $\int_0^\infty dt_3\int_{0}^{t_3}dt$ for the first term, that can be reordered as $\int_0^{\infty}dt\int_{t}^{\infty}dt_3$, and upon reintroducing the step function one obtains
	\begin{equation}
	\Pi^1(\omega)=P^0_{n_1n_2}(\omega)\widetilde{K}_{\scriptsize\begin{array}{l}n_1n_2\\n_3n_4\end{array}}(\omega)P^0_{n_3n_4}(\omega),
	\end{equation}
	with
	\begin{eqnarray}
	\widetilde{K}_{\scriptsize\begin{array}{l}n_1n_2\\n_3n_4\end{array}}(\omega)=\int_0^\infty dt e^{\pm i(\omega+\varepsilon_{n_2}-\varepsilon_{n_3})t}K_{\scriptsize\begin{array}{l}n_1n_2\\n_3n_4\end{array}}(\pm t)+ 
	\int_0^\infty dt e^{\pm i(\omega+\varepsilon_{n_4}-\varepsilon_{n_1})t}K_{\scriptsize\begin{array}{l}n_1n_2\\n_3n_4\end{array}}(\mp t),
	\end{eqnarray}
where the upper(lower) sign is for $n_1n_2,n_3n_4=ph,ph(hp,hp)$. Comparing to the work of Marini and Del Sole\cite{PhysRevLett.91.176402}, the kernel has two terms, corresponding to the two time orderings of the interaction and is expressed as the Laplace transform ($\int_0^{\infty}e^{-st}f(t)dt$) calculated at $\mp i(\omega+\varepsilon_{n_2}-\varepsilon_{n_3})$ and $\mp i(\omega+\varepsilon_{n_4}-\varepsilon_{n_1})$, and for direct comparison replace $n_1n_2,n_3n_4$ with $v_1c_1,v_2c_2$ or $c_1v_1,c_2v_2$. It is important to note that for the time reversed part of the interaction (i.e., for a screened interaction that propagates backwards in time), the integral is for $t<0$ (since the backwards propagating component is zero for $t>0$), but this integral is transformed to positive times, which will now require calculating the Laplace transform of $K(-t)$.
The kernel can be written as a combination of an instantanteous and a forward and backward propagating interaction, as shown for the screening above and will be discussed below for the other terms that appear in the expression for the vertex. Focusing only on the screening in the vertex for now (the term that gives rise to the usual BSE, but this time with dynamical effects) produces the kernel\cite{cunningham24}
\begin{eqnarray}
K_{\scriptsize\begin{array}{l}n_1n_2\\n_3n_4\end{array}}(t)=v_{\scriptsize\begin{array}{l}n_1n_2\\n_3n_4\end{array}}\delta(t)-v_{\scriptsize\begin{array}{l}n_4n_2\\n_3n_1\end{array}}\delta(t)-\sum_{\scriptsize\begin{array}{l}n_5n_6\\n_7n_8\end{array}}v_{\scriptsize\begin{array}{l}n_4n_2\\n_5n_6\end{array}}\Pi_{\scriptsize\begin{array}{l}n_5n_6\\n_7n_8\end{array}}(-t)v_{\scriptsize\begin{array}{l}n_7n_8\\n_3n_1\end{array}},
\end{eqnarray}
that in the frequency domain produces the first line from Eq.~\ref{eq:omegaK} below.

\begin{figure*}[t]
\includegraphics[width=0.95\textwidth,clip=true,trim=0.3cm 7.0cm 0.0cm 0.0cm]{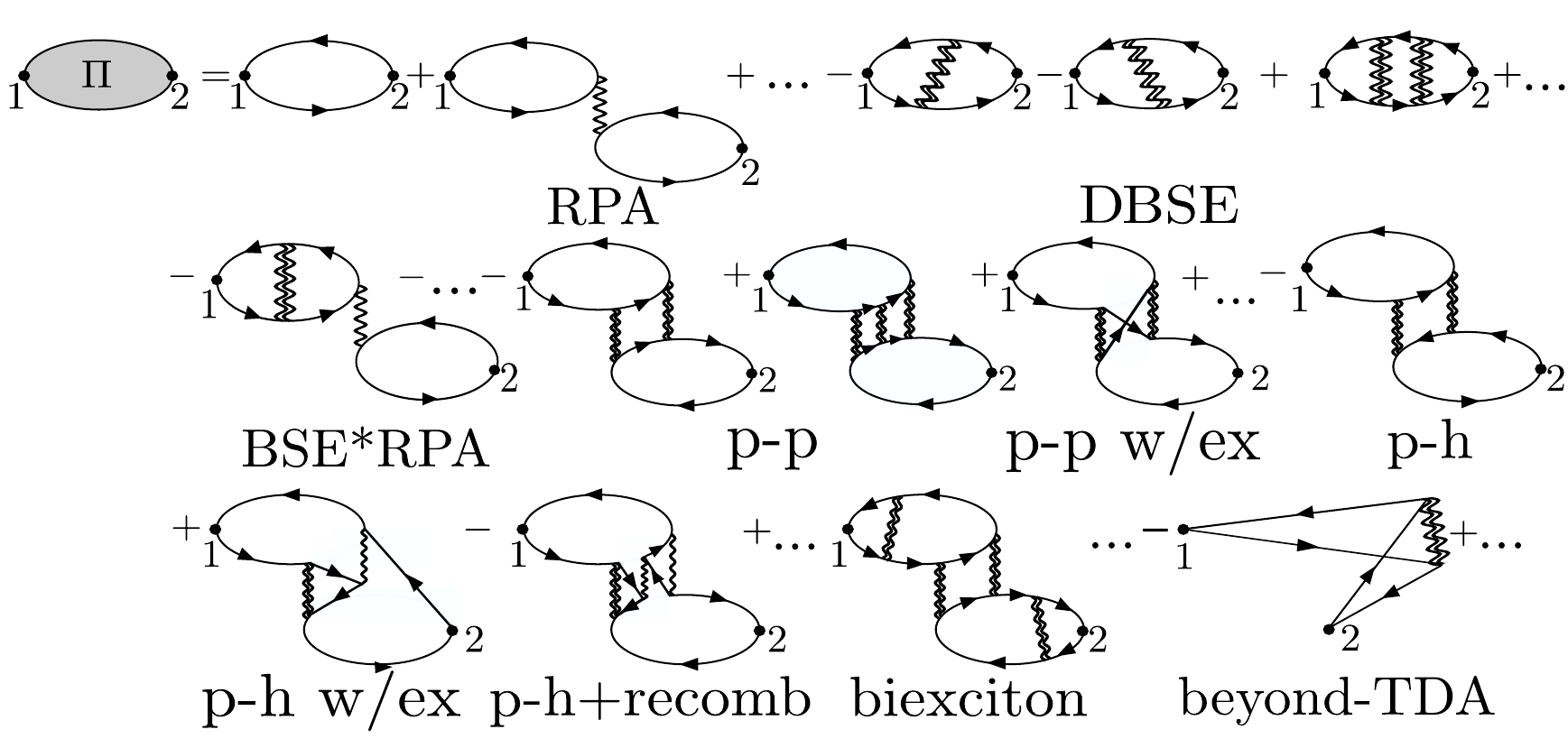}
\caption{A subset of the diagrams included in the method for the reducible polarization function. The first two terms are part of an infinte series of bubbles that produce the RPA, followed by the infinite ladder series of time-dependent interactions within these bubbles (the dynamical BSE) that excites further electron-hole pairs. Then is the creation of an electron hole pair by the excited particle that arises when considering $\delta W/\delta G$ in the interaction kernel (with similar diagrams for the hole) and the infinte series of interactions between the electrons and holes with a direct and exchange contribution (from considering $\delta\Gamma/\delta G$ in the interaction kernel). These channels are coupled, as demonstrated by, e.g., BSE*RPA, p-p with exchange and the multiexcitonic term. The last term is the coupling diagram for the BSE beyond the TDA (see text). In these diagrams, time can be thought to flow horizontally. Straight lines represent propagators (Green's functions) and single (double) wavy lines represent the bare(screened) Coulomb interaction. The $GW$ self-energy is determined by adding two vertical (instantaneous) Coulomb interaction lines; one that attaches at one end to 1 (the other interaction to 2) and the other ends connect to the ends of a propagator describing the added particle(hole).}
\label{fig:diagramsPi}
\end{figure*}

Upon including higher order terms in the expansion of $\Pi$, i.e., $\Pi^{(2)}=P^0KP^0KP^0$ it can be seen that a ladder series of interactions of this form is present in the reducible polarization function $\Pi(\omega)=P^0(\omega)+P^0(\omega)\widetilde{K}(\omega)\Pi(\omega)$. The kernel for $n_1n_2$ and now $n_4n_3$ particle-hole(hole-particle) pairs (i.e., going beyond the TDA and including the coupling between positive and negative energy transitions; discussed further below) is
\begin{eqnarray}
	\widetilde{K}_{\scriptsize\begin{array}{l}n_1n_2\\n_3n_4\end{array}}(\omega)=\int_0^\infty dt e^{\pm i(\varepsilon_{n_3}-\varepsilon_{n_1})t}K_{\scriptsize\begin{array}{l}n_1n_2\\n_3n_4\end{array}}(\mp t)+ 	\int_0^\infty dt e^{\pm i(\varepsilon_{n_2}-\varepsilon_{n_4})t}K_{\scriptsize\begin{array}{l}n_1n_2\\n_3n_4\end{array}}(\pm t),
	\end{eqnarray}
where the upper(lower) sign is for $n_1n_2$ a particle-hole(hole-particle) pair and $n_3n_4$ a hole-particle(particle-hole) pair. The coupling between positive and negative energy transitions can be understood by considering the BSE in the time domain. Even for an instantaneous interaction kernel, the interaction can occur at a time before or after the external particle or hole (that polarizes the system) is added, see Fig.~\ref{fig:diagramsPi}. Since the interaction is between positive and negative energy propagating pairs (travelling forwards and backwards in time, respectively) and not restricted to the interval where the external particle is present, the interaction is not shifted by the perturbing $\omega$. For time-dependent interactions we can also consider the case of $n_1$ and $n_2$ both propagating forward or backward in time; considering $P^0_{n_1n_2}(1;3,4)=-iG_{n_1}(1,3)G_{n_2}(4,1)$ for $t_3<t_1$ and $t_4>t_1$. These processes describe complex quantum mechanical phenomena such as the emergence of virtual pairs that exist temporarily describing quantum/vacuum fluctuations (as a result of the uncertainty principle) and/or representing intermediate states mediated by the presence of the time-dependent screening. This is beyond the scope of this work, but worth mentioning in the hope of stimulating curiousity.

The functional derivative $i\delta\Sigma/\delta G$ is discussed in detail in Ref.~\cite{cunningham24}, where it is included in the expression for the self-energy $\Sigma=iGW\Gamma$. This functional derivative, and the resulting vertex from that work, will now be used to derive an expression for the kernel in the BSE for the reducible polarization function. In that work, three body correlated propagation was suppressed (e.g., diagrams that include correlations between the added particle and both the excited electrons and the holes). The diagrams (for the self-energy) that are neglected in that work are presented in figures 1 and 2 in Ref.~\cite{cunningham24} and the corresponding diagrams for the polarization are thus straightforward to deduce. From Eq.~15 in that work, the functional derivative can be written as
\begin{align}i\frac{\delta\Sigma(3,4)}{\delta G(5,6)}=-W(4,3)\delta(3,5)\delta(4,6)-L_{pp}(3,6;7,8)W(6,3)\chi(7,4;5,8)-L_{ph}(3,5;7,8)W(5,3)\chi(7,4;8,6),\label{eq:dSigdGL}
\end{align}
where it was assumed that all appearances of $W$ within the functional derivative are instantaneous (which will also be assumed in this work, except for the first term that is allowed explicit dependence on time to include dynamical excitonic effects) and it is also assumed that the vertices that are attached to the $W$ in this recursive series for the vertex (Eq.~14 from Ref.~\cite{cunningham24}) are approximated as $\Gamma(1,2;3)=\delta(1,2)\delta(2,3)$\footnote{It is discussesed in Ref.~\cite{cunningham24} how this can be a good approximation with the correct use of non-interacting Green's functions as a result of the Ward identity, and this will be discussed further in the conclusions.} and $L$ are two-body correlated propagators that contain correlations between the particles and holes (series of ladder interactions) in the system,
\begin{align}
L_{pp}(1,2;3,4)=&-iG(1,3)G(2,4)+\left\{-iG(1,5)G(2,6)\right\}\times\left\{-W(6,5)\right\}L_{pp}(5,6;3,4)\nonumber\\
L_{ph}(1,2;3,4)=&-iG(1,3)G(4,2)+\left\{-iG(1,5)G(6,2)\right\}\times\left\{-W(6,5)\right\}L_{ph}(5,6;3,4)\label{eq:L}
\end{align}
and
 \begin{align}
\chi(7,4;5,8)=&W(4,5)\delta(7,4)\delta(5,8)-W(4,7)\delta(7,5)\delta(4,8)\label{eq:chi},
\end{align}
that contain a direct and an exchangelike contribution, accounting for exchange between electrons/holes from different electron-hole pairs (arising from the first term on the right hand side of Eq.~\ref{eq:dSigdGL} when considering the recursive series for the vertex). In this work, 3-body correlated propagation is suppressed.\cite{cunningham24} $L_{ph}(1,2;3,4)$ is the four-point extension of the usual particle-hole propagator, i.e., the irreducible polarization function $P$. It was shown in Ref.~\cite{cunningham24} by including higher order vertices that $L_{ph}$ can be replaced with the reducible polarization function $\Pi$, which effectively describes recombination with the creation of new electron-hole pairs, as in the RPA.
The reducible polarization function is then of the form of the BSE, with kernel
\begin{align}
K(3,4;5,6)=&v(3,5)\delta(3,4)\delta(5,6)\delta(t_3-t_5)-W(4,3)\delta(3,5)\delta(4,6)-\nonumber\\&L_{pp}(3,6;7,8)W(6,3)\chi(7,4;5,8)-
\Pi(3,5;7,8)W(5,3)\widetilde{\chi}^{ph}(7,4;8,6),\label{eq:Ktime}
\end{align}
where $\widetilde{\chi}^{ph}$ is the interaction in Eq.~\ref{eq:chi} but with the exchange interaction replaced with a bare interaction (discussed in Fig.~5 and the surrounding text in Ref.~\cite{cunningham24}). 

The particle-particle and hole-hole, non-correlated propagators $L(1,2,3,4)=-iG(1,3)G(2,4)$ for a single time difference are $\mp 1/(\omega-(\varepsilon_{n_1}+\varepsilon_{n_2})\pm i\eta)$ in frequency, or $ie^{-i(\varepsilon_{n_1}+\varepsilon_{n_2})t}e^{\mp\eta t}\theta(\pm t)$ in time, and the kernel is then
\begin{eqnarray}
K_{\scriptsize\begin{array}{l}n_1n_2\\n_3n_4\end{array}}(t)=v_{\scriptsize\begin{array}{l}n_1n_2\\n_3n_4\end{array}}\delta(t)-W_{\scriptsize\begin{array}{l}n_4n_2\\n_3n_1\end{array}}(-t)-{\displaystyle\sum_{\scriptsize\begin{array}{l}n_5n_6\\n_7n_8\end{array}}}L_{\scriptsize\begin{array}{l}n_5n_6\\n_7n_8\end{array}}(t)\tilde{W}_{\scriptsize\begin{array}{l}n_4n_6\\n_5n_1\end{array}}\chi_{\scriptsize\begin{array}{l}n_7n_2\\n_3n_8\end{array}}-{\displaystyle\sum_{\scriptsize\begin{array}{l}n_5n_6\\n_7n_8\end{array}}}\Pi_{\scriptsize\begin{array}{l}n_5n_6\\n_7n_8\end{array}}(t)\tilde{W}_{\scriptsize\begin{array}{l}n_6n_3\\n_5n_1\end{array}}\tilde{\chi}_{\scriptsize\begin{array}{l}n_7n_2\\n_8n_4\end{array}}^{ph},
\end{eqnarray} 
where the $\tilde{W}$ are assumed static\footnote{The method outlined in this work could be extended to describe time-dependent interactions in the $T$-matrix diagrams that result from including the $L$ in Eq.~\ref{eq:dSigdGL} from Ref.~\cite{cunningham24}.} (see Fig.~\ref{fig:diagramsPi}). The reducible polarization function with this kernel is then represented in Fig.~\ref{fig:diagramsPi} and in the frequency domain the kernel is 
\begin{widetext}\small{
\begin{align}
\widetilde{K}_{\scriptsize\begin{array}{l}n_1n_2\\n_3n_4\end{array}}(\omega)=&v_{\scriptsize\begin{array}{l}n_1n_2\\n_3n_4\end{array}}-v_{\scriptsize\begin{array}{l}n_4n_2\\n_3n_1\end{array}}\mp\displaystyle\sum_{\lambda,\scriptsize\begin{array}{l}n_5n_6\\ph\end{array}}v_{\scriptsize\begin{array}{l}n_4n_2\\n_5n_6\end{array}}X_{n_5n_6,\lambda}\left\{\displaystyle\frac{X_{\lambda,hp(ph)}^{-1}v_{\scriptsize\begin{array}{l}hp(ph)\\n_3n_1\end{array}}}{\omega+\varepsilon_{n_2}-\varepsilon_{n_3}+E_\lambda\pm i\eta}+\frac{X_{\lambda,ph(hp)}^{-1}v_{\scriptsize\begin{array}{l}ph(hp)\\n_3n_1\end{array}}}{\omega+\varepsilon_{n_4}-\varepsilon_{n_1}-E_\lambda\pm i\eta}\right\}\nonumber\\&
\mp\displaystyle\sum_{\alpha,\scriptsize\begin{array}{l}pp'(hh')\\n_7n_8\end{array}}\left\{\tilde{W}_{\scriptsize\begin{array}{l}n_4p'(h')\\p(h)n_1\end{array}}\displaystyle\frac{Y_{pp'(hh'),\alpha} Y_{\alpha,n_7n_8}^{-1}}{\omega+\varepsilon_{n_2}-\varepsilon_{n_3}-\Omega_\alpha\pm i\eta}
+\displaystyle\sum_{\alpha,\scriptsize\begin{array}{l}hh'(pp')\\n_7n_8\end{array}}\tilde{W}_{\scriptsize\begin{array}{l}n_4h'(p')\\h(p)n_1\end{array}}\displaystyle\frac{Y_{hh'(pp'),\alpha} Y_{\alpha,n_7n_8}^{-1}}{\omega+\varepsilon_{n_4}-\varepsilon_{n_1}+\Omega_\alpha\pm i\eta}\right\}\chi_{\scriptsize\begin{array}{l}n_7n_2\\n_3n_8\end{array}}\nonumber\\&\pm\displaystyle\sum_{\lambda,\scriptsize\begin{array}{l}n_5n_6\\ph\end{array}}\tilde{W}_{\scriptsize\begin{array}{l}n_6n_3\\n_5n_1\end{array}}X_{n_5n_6,\lambda}\left\{\displaystyle\frac{X_{\lambda,ph(hp)}^{-1}\tilde{\chi}^{ph}_{\scriptsize\begin{array}{l}p(h)n_2\\h(p)n_4\end{array}}}{\omega+\varepsilon_{n_2}-\varepsilon_{n_3}-E_{\lambda}\pm i\eta}+\frac{X_{\lambda,hp(ph)}^{-1}\tilde{\chi}^{ph}_{\scriptsize\begin{array}{l}h(p)n_2\\p(h)n_4\end{array}}}{\omega+\varepsilon_{n_4}-\varepsilon_{n_1}+E_{\lambda}\pm i\eta}\right\},\label{eq:omegaK}
\end{align}}
\end{widetext}
where the upper(lower) sign is for $n_1$ a particle and $Y$ and $\Omega$ are the eigenvalues and eigenvectors of the particle-particle (hole-hole) matrix, that in the particle-particle basis is $(\varepsilon_{n_1}+\varepsilon_{n_2})\delta_{n_1n_3}\delta_{n_2n_4}\pm W_{\scriptsize\begin{array}{l}n_2n_4\\n_3n_1\end{array}}$. Note that for convenience the beyond-TDA terms have been omitted. Although this equation appears challenging, it is constructed in a similar way to the usual static BSE. The third line is constructed with the same elements as in the first line; and the second line is constructed analogously, instead using the particle-particle(hole-hole) eigenvectors and eigenvalues (requiring the diagonalization of a particle-particle and hole-hole matrix, analogous to the particle-hole matrix) that can also be used to construct the $T$-matrix contributions to the self-energy in Ref.~\cite{cunningham24}, describing correlations between the added particle(hole) and the excited electrons(holes). As mentioned above, a reduced, mixed particle basis can be used to reduce the memory/storage demands, and the matrix multiplications can be handled with modern accelerated computing techniques. Once the frequency-dependent kernel is constructed, the usual BSE for the reducible polarization function can be solved by either inversion or diagonalization for the particular frequency. The actual size of the kernel is identical to conventional BSE implementations, such as those in Refs~\cite{BSE_paper,PhysRevLett.91.056402,PhysRevLett.91.256402}, however, the frequency dependence in the kernel would now require the diagonalization (or inversion) of the 2-particle matrix for a range of frequencies, since the kernel is now frequency dependent. Since the reducible polarization function is such that we know the poles of $\Pi$ exactly, the convolution between $G$ and $\Pi$ (to produce the $GW$+second-order screened exchange self-energy) can be performed analytically, making use of the Cauchy residue theorem, and so the screening only needs to be calculated with $\omega$ determined from the poles of $G$ and the chosen input energy $\omega$, that is often set equal to the single particle eigenvalues in many methods, including many QS$GW$ implementations.\cite{QSGW_prl1,QSGW_PRL,QSGW_paper} Note that the indices are composite indices containing wavevector $\boldsymbol{k}$ in extended systems and their combinations are dictated by the conservation of momentum. 

\section{Conclusions}
An expression for the kernel in the BSE for the reducible polarization function was derived from Hedin's equations. This includes a number of 2-body interactions beyond the usual BSE as a result of now including the functional derivative of the screening and vertex in $i\delta\Sigma/\delta G$, {\textit{abinitio}}. These vertex contributions can be included in the kernel after considering their contribution to the self-energy (giving rise to second-order screened exchange and $T$-matrix channel contributions\cite{cunningham24}). A frequency-dependent interaction kernel incorporating RPA, dynamical excitonic and multiexcitonic effects - due to correlations between the particles and holes from different electron-hole pairs - is presented. The terms that couple the positive and negative energy transitions (i.e., going beyond the Tamm-Dancoff approximation) are also considered and found to be independent of frequency and this is discussed along with an intuitive description of the physical processes described by these terms. In order to present a tractable equation, a number of approximations were made. Three-body correlated propagation was suppressed; the vertices that attached to the screening within the functional derivative $i\delta\Sigma/\delta G$ were approximated as a product of Dirac delta functions (as discussed above) and beyond the first term in $i\delta\Sigma/\delta G$ (the term that appears in the standard BSE), any appearance of $W$ is assumed static. In order to construct the kernel, the eigenvectors and eigenvalues of the usual static BSE had to be determined, along with the eigenvectors and eigenvalues of the particle-particle(hole-hole) matrix that produce the $T$-matrix diagrams in the self-energy.\cite{cunningham24}\par
The result is the usual 2-body BSE matrix that requires inversion (or diagonalization), but now with a frequency-dependent interaction kernel containing the added diagrams. Since the reducible polarization function is calculated instead of the irreducible one, this matrix only needs inverted at a range of known energies (as opposed to requiring the value at a dense range of frequencies required for the numerical frequency convolution to produce the self-energy when using the irreduciible one\cite{QSGW_paper}). The reducible polarization function also has the benefit that macroscopic quantities such as electron energy loss spectrospcopy (EELS), $1+v\Pi$, and optical absorption spectra, $1/(1+v\Pi)$,  are readily computed.  
\par As is the case with the standard BSE, the number of states that can be included will be much less than the total number of possible states in many systems of interest, however, states not included in the method can still be included at a lower level of theory, e.g., RPA, as discussed in Refs~\cite{cunningham24,QSGWhat}. 
Self-polarization/correlation effects (significant in strongly correlated systems) will be present in the method, however, these are reduced by the competing direct and exchange interactions and can be mitigated by taking steps to ensure states are not included more than once in the expression for the polarization function and self-energy (another benefit of making use of the reducible polarization function over the irreducible one). The methods can be adopted to consider time-dependent interactions in the $T$-matrix diagrams and to also include second-order dynamically screened exchange effects. The method is general and can be incorporated into, for e.g., the quasiparticle self-consistent $GW$ formalism\cite{QSGW_paper,QSGWhat}; that the author has previously extended by incorporating excitonic effects with only the static part of $W$\cite{QSGWhat}.\par
As discussed extensively in Ref.~\cite{cunningham24}, the use of non-interacting Green's functions in many-body theory tends to rely on a cancellation with the vertex, dictated by the Ward identity\cite{QSGW_paper,PhysRevA.90.032506}. However, since vertices are now being treated explicitly, it may become necessary to abandon the quasiparticle approximation and consider interacting Green's functions (as considered in Ref.~\cite{QSGWhat}) and/or make use of so-called $Z$-factors (investigated when examining time-dependent excitonic effects in Ref.~\cite{PhysRevLett.91.176402}). With the additions discussed above, along with the possibility of including 3-body correlation effects,\cite{stef,PhysRevLett.131.216401} and the inclusion of spin fluctuations\cite{Stepanov19} (where $T$-matrix channels\cite{Friedrich21} and the particle-particle correlation function that governs superconductivity\cite{Acharyalafeas} are calculated) and phonon effects,\cite{savio24} a broadly applicable, high-fidelity {\textit{ab~initio}} approach to solving one-, two-, and even three-particle properties of the many-body problem is within reach. The formalism presented will prove beneficial for researchers studying electronic structure theory, nonlinear optics and high-harmonic generation in general systems of interest, including strongly correlated materials.
\begin{acknowledgments}
The author would like to thank Dermot Green (QUB) for fruitful discussions and all those involved in the {\it CCP flagship project: Quasiparticle Self-Consistent $GW$ for Next-Generation Electronic Structure}, especially Mark van Schilfgaarde (NREL, Golden, Co.) and Myrta Gr\"{u}ning (QUB). The data that support the findings of this study are available from the corresponding author upon reasonable request. The data consist of Feynman diagrams and analytical expressions presented within the manuscript; no numerical or experimental data were generated.
\end{acknowledgments}
\bibliography{references}
\end{document}